# Ideal type-II Weyl phonons in wurtzite CuI

Jian Liu,[1,2,§] Wenjie Hou,[3,4,§] En Wang,[1,2] Shengjie Zhang,[1,2] Jia-Tao Sun*[1,2] and Sheng Meng*[1,2,5]

[1]Beijing National Laboratory for Condensed Matter Physics and Institute of Physics, Chinese Academy of Sciences, Beijing, 100190, P. R. China

[2]School of Physical Sciences, University of Chinese Academy of Sciences, Beijing 100190, China

[3]Fert Beijing Institute, BDBC, School of Microelectronics, Beihang University, Beijing 100191, China

[4]Beihang-Goertek Joint Microelectronics Institute, Qingdao Research Institute, Beihang University, Qingdao 266104, China

[5]Songshan Lake Materials Laboratory, Dongguan, Guangdong, 523808, P. R. China

[§]These authors contributed equally to this work.
*Email: jtsun@iphy.ac.cn, smeng@iphy.ac.cn

**Abstract:** Weyl materials exhibiting topologically nontrivial touching points in band dispersion pave the way to exotic transport phenomena and novel electronic devices. Here, we demonstrate the signature of ideal type-II Weyl phase in phonon dispersion of solids through first-principles investigations. Type-II phononic Weyl phase is manifested in noncentrosymmetric wurtzite CuI by six pairs of Weyl points (WPs) in the $k_z = 0.0$ plane. On the iodine-terminated surface of the crystal, very clean surface arcs are readily detectable. Each pair of WPs connected by open surface arcs is well separated by a large distance of 0.26 Å$^{-1}$, ten times larger than that in TaAs. The opposite chirality of WPs with quantized Berry curvature produces Weyl phonon Hall effect, in analogy to valley Hall effect of electrons. Such ideal type-II Weyl phase are readily observable in experiment, providing a unique platform to study novel thermal transport properties distinct from type-I Weyl phase.

Since the theoretical prediction and experimental realization of Weyl semimetal state in the TaAs class of compounds [1-4], Weyl materials including Weyl fermions [5-12], Weyl photonic crystal [13,14] and Weyl phononic crystal [15-20] with novel surface states have attracted significant interest recently. The spin-1/2 Weyl points (WPs) emerging in solids without time reversal symmetry or inversion symmetry are the twofold degenerated crossing points of two linearly dispersing bands in the three directions of momentum space [21,22]. The quantized Berry curvature enclosing the WPs in momentum space can be viewed as magnetic monopole of opposite chirality, leading to the topologically protected surface (Fermi) arcs [23,24]. Compared to the type-I WP [13,25-27], the type-II WP phase is expected to realize very distinct transport properties, such as anisotropic chiral anomaly [28], anomalous Hall effect [29-31], enhanced thermoelectric coefficient [32,33]. Because of the strongly tilted electron and hole pockets in type-II Weyl semimetals, the type-II Weyl cone easily induces the overlapping states between the Weyl cone and the trivial bulk bands [25]. The overlapping states can induce the Weyl cone with different energy and different momentum (Fig. 1(a)). To date, the ideal type-II Weyl phase without the overlapping states have not been reported.

In fact, if the Weyl phase is ideal, the surface states show open surface arcs [13]; Otherwise, at least one of the WP will be wrapped up in the trivial bulk pockets and thereby the surface states show closed surface arcs, which blurs the observation of topological nontrivial surface states in experiment. We first demonstrate what the ideal type-II Weyl phase looks like. As shown in Fig. 1, a pair of pocket can form two WPs if the two bands cross each other, or only one WP if the two bands simply touch at a given point. For the former case (Fig. 1(a)), one WP at least will be unavoidably wrapped up in the trivial bulk pockets. Most discovered type-II Weyl materials belong to this category [34-42]. Nevertheless, for the latter case, all WPs are related to each other by symmetry, and thereby locate at the same energy. In particular, when the bulk bands away from the Weyl cone have an observable gap ($E_g > 0$) (Fig. 1(c)), the strong hybridization between the Weyl cone and the bulk bands can be minimized. The corresponding Weyl state is called ideal type-II Weyl phase. Though some realistic candidates [9,19,43] meet the requirement mentioned in Fig. 1(b), they do not have the observable gap ($E_g < 0$) and then are not ideal. Thus, finding the ideal type-II Weyl phase with clean surface states remain a challenging task.

Here, we show by the first-principles calculations that the wurtzite CuI features the ideal type-II Weyl phonon, exhibiting six pairs of accidentally degenerated WPs at the frequency of 3.60 THz in the $k_z = 0.0$ plane. The ideal feature of the type-II Weyl phonon phase is manifested by vanishing band overlapping with the bulk phonon continuum, which leads to clean surface arcs. The separation for each pair of WP is 0.26 Å$^{-1}$, ten times larger than that of TaAs solids, suggesting that wurtzite CuI is a promising candidate for studying the spin-1/2 type-II Weyl phonon. The opposite Berry curvature for WPs in the Weyl phonon can host the phonon Hall effect under the excitation of polarized photon, in analogy to the valley Hall effect for fermions.

*Phonon dispersion and Weyl points.* As shown in Fig. 2(a), wurtzite cuprous iodide CuI exists at high temperature (370-400 ℃) and has a hexagonal lattice in a noncentrosymmetric space group of $P6_3mc$ (No. 186) [44,45]. It is isomorphic to the point group $C_{6v}$ respecting the mirror symmetries $\sigma_v(-k_x, k_y)$, six-fold rotational symmetry $C_{6z} = (\frac{k_x}{2} - \frac{\sqrt{3}k_y}{2}, \frac{\sqrt{3}k_x}{2} + \frac{k_y}{2}, k_z)$ and time reversal symmetry $\mathcal{T}$. The phonon dispersion of wurtzite CuI along high-symmetry path in the whole Brillouin zone (BZ) is shown in Fig. 2(c). The absence of vibration modes of negative frequency at the whole BZ indicates the strong lattice stability of CuI. One can find that a tiny gap appears along the Γ−M and Γ−K direction around the optical phonon frequency of 3.60 THz, indicating the possibility of the crossing points around. To confirm that these two optical branches indeed touch, we plot the contour of the frequency gap between the sixth and seventh optical branches in the $k_z = 0.0$ plane of the whole BZ by applying a much dense grid [46] of momentum points in Fig. 3(a). The frequency gap approaches zero at the point $(0.1866, 0.7197, 0.0)$ Å$^{-1}$, located slightly away from any high-symmetry momentum path. The positions of other crossing points are associated by mirror symmetries $\sigma_v$, rotational symmetry $C_{6z}$, as shown schematically in Fig. 2(b).

As phonons do not obey the Pauli exclusion principle [18,47], the phonon properties can be studied in the whole range of phonon frequencies. In this letter, we focus on the crossing point of the sixth and seventh optical branches at the frequency of 3.60 THz. Previous results in electronic system show accidental degeneracy in three-dimensional space can lead to the appearance of WP at generic points, though no corresponding symmetry protects the gapless points [22].

*Topology of Weyl phonon.* To identify the topological properties of these WPs, the Berry curvature is calculated via $\Omega_n^z(k) = \nabla_k \times \langle u_{nk}|i\nabla_k|u_{nk}\rangle$, where $n$ labels the band number and $|u_{nk}\rangle$ is the phonon eigenmodes [22,42]. Fig. 3(b) shows the monopole-like distribution of the Berry curvature around two crossing nodes in the $k_z = 0.0$ plane. We have also checked the chiral charge of the crossing nodes by integrating the Berry curvature calculated on a closed surface enclosing the corresponding WP. We found that their Chern number $\mathcal{C}$ is +1 for source (blue) and −1 for sink (red), respectively. Thus, the twelve crossing nodes in the $k_z = 0$ plane are indeed WPs [22].

In order to understand the WPs in CuI, we show an enlarged view of the phonon dispersion around a WP in Fig. 2(d). The corresponding 3D plot of the crossing branches with $k_z = 0.0$ is displayed in Fig. 2(e). It is clear that the two bands cross with linear dispersion with homochromous velocity along one direction and the Weyl cone is strongly tilted, indicating the identified twelve WPs belong to the type-II WPs [25]. Furthermore, compared to the tiny gap in the $k_z = 0.0$ plane, the gap out of the $k_z = 0.0$ plane between the sixth and seventh optical branches is much larger, as shown in Fig. 2(a). Therefore, we identify the CuI to be an ideal type-II Weyl phononic materials based on our definition of ideal type-II WPs mentioned above.

Once the crossing nodes are identified as WPs, we can utilize symmetries, $\sigma_V$ and $C_{6z}$ to understand the distribution of the WPs in momentum space [48]. Suppose a WP with chirality $\mathcal{C} = +1$ exists at a generic position $(k_x, k_y, 0.0)$, the other eleven WPs can be obtained by above symmetry operations. We find any high-symmetry lines can not host WPs: (i) M−K line. Setting $k_y$ to be 0.5, the operation $\sigma_v$ and $C_{6z}^3$ maps the WP $(k_x, 0.5, 0.0)$ with chirality $\mathcal{C} = +1$ to be $(-k_x, 0.5, 0.0)$ with $\mathcal{C} = -1$ and $(-k_x, -0.5, 0.0)$ with $\mathcal{C} = +1$, respectively. The two new WPs do not meet the crystal periodicity condition. (ii) Γ−M line. Taking $k_x$ to be zero, the operation $C_{6z}$ and $C_{6z}^4 \sigma_v C_{6z}^3$ maps the WP $(0.0, k_y, 0.0)$ to the same position $(-\frac{\sqrt{3}k_y}{2}, \frac{k_y}{2}, 0.0)$, but with the opposite chirality. (iii) Γ−K line. Using the relation $k_y = \sqrt{3} k_x$, the operation $C_{6z}^5$ and $C_{6z} \sigma_v C_{6z}^3$ maps the WP $(k_x, \sqrt{3}k_x, 0.0)$ to the same position $(2k_x, 0.0, 0.0)$, but with the opposite chirality. Indeed, WPs will annihilate in pairs if two WPs with the opposite chirality meet in the momentum space [22]. We find that the discussion above that the WPs can not locate at any high-symmetry lines is consistent with the previous first-principles calculations [9].

*Three band effective Hamiltonian.* In order to qualitatively understand the existence of the Weyl phase, we construct a three band effective $\boldsymbol{k} \cdot \boldsymbol{p}$ Hamiltonian by applying the symmetry principle of Γ point based on the theory of invariants [49]. Similar to the case in SrHgPb electronic system [9], we find the obtained Hamiltonian can describe the essential physics of the Weyl phase away from the Γ point. The effective Hamiltonian near the Γ point is dictated by the lattice symmetry ($C_{6v}$) and the time reversal symmetry. Our first-principles calculations show that the upper and lower part of the Weyl cone belong to the two-dimensional representation $E_1$ and one-dimensional representation $B_1$, respectively, as shown in Fig. 2(a). It is natural to consider three bands as the basis in the effective Hamiltonian. As phonons do not have the concept of the electronic orbital, we only require that the form of the real basis can describe the corresponding irreducible representation [18]. Thus, the basis at the Γ point can be written as $\{x, y\}$ for $E_1$ representation and $\{y^3 - 3yx^2\}$ for $B_1$ representation. Considering the out-of-plane bands form an obvious gap, we confine the effective Hamiltonian in the $k_z = 0.0$ plane. Thus, the most general effective Hamiltonian up to $O(k^4)$ near the Γ point is given by:

$$H_{eff}(k) = \varepsilon(k) + A(k)\left[(k_x^2 - k_y^2)I_y - k_x k_y I_z\right] + B(k)\left[k_x k_y I_{xy} - \right.$$

$$\left(k_x^2 - k_y^2\right)I_{zx}\right] + C(k)I_x^2 + H_w \tag{1}$$

where $\varepsilon(k) = D_1 + D_2 k_\parallel^2$, $A(k) = A_1 + A_2 k_\parallel^2$, $B(k) = B_1 + B_2 k_\parallel^2$, $C(k) = C_1 + C_2 k_\parallel^2$ and $k_\parallel^2 = k_x^2 + k_y^2$. Combinations of $\{1, k_x^2 - k_y^2, k_x k_y\}$ and $\{I_{yz}, I_y^2, I_z^2\}$ will generate the $H_w$ term (part IV Supplemental Material). $I_y$, $I_z$, $I_{xy}$, $I_{zx}$, $I_{yz}$, $I_x^2$, $I_y^2$ and $I_z^2$ are the matrix basis. The various constant $A_i$, $B_i$, , $C_i$, $D_i$, $E_{ij}$ and $F_{ij}$ describe the specific band properties.

By fitting the phonon bands of the effective model with the first-principles calculations around the Γ point, the parameters in the three band effective model are determined and are summarized in Table S1. The fitted phonon dispersions for wurtzite CuI are plotted in Fig. S6. It is shown that the phonon dispersions at WP in the $k_z = 0.0$ plane generally agree well with the first-principles calculations. Hence, the effective model in Eq. 1 can capture the topological properties of the Weyl phase, confirming the existence of the Weyl phase in CuI. We find from the effective model that not the Weyl phase but the node-line phase will be generated when the term $H_w$ vanishes.

*Topological surface states of Weyl phonon.* Similar to the electronic system, the existence of Weyl phonons should also give rise to the topologically protected nontrivial surface arcs in momentum space connecting the WPs of opposite chirality [18,23]. Generally, the coupling within the first surface layer of polar semiconducting CuI is different from the bulk. The change in the surface force constants can lead to the shift of the surface phonon frequency in polar semiconductors within 15% making our prediction still reasonable [50,51]. As shown in Fig. 4, we calculated the surface phonon states and the surface spectral function for a fixed frequency $f = 3.60$ THz on the CuI(0001) (Cu-terminated) surface and the CuI($000\bar{1}$) (I-terminated) surface. The blue and red spheres denote the WPs with chirality of $\mathcal{C} = +1$ and $\mathcal{C} = -1$, respectively. The white regions represent the projected bulk bands, whereas the red color lines show the surface states. As expected, the surface arcs and surface states connecting a pair of WPs with opposite chirality are clearly observable. On the

Cu-terminated surface, as shown in Figs. 4(a)-(c), we can identify two circled dispersing states around the WPs: states $\beta$ from bulk pocket bands form a closed path and thus contribute to the trivial arcs; states α connect a pair of WPs with opposite chirality and thus are the candidate topological arcs [42]. On the I-terminated surface (Figs. 4(d)-(f)), very clean surface arcs are observed. Because of the broken inversion symmetry of the structure, the connecting pattern of surface arcs changes but does not disappear [46].

Unlike the most type-II Weyl semimetals where the strong hybridization of bulk and surface states makes the identification of topological character of Fermi or surface arcs very hard, the surface arcs in CuI phonons are very clean. The differences lie in the fact: (i) the twelve type-II WPs in CuI locate at the same frequency with the same shape of phonon dispersion determined by the crystal symmetry and time reversal symmetry, while two type-II WPs in the former case with opposite chirality have a shifted energy or frequency; (ii) the pockets away from the WPs are well separated and form an observable gap. In contrast, it is noted that the pockets which form type-II WPs in the phonon dispersion of TiS, ZrSe and HfTe overlap too much [19], making the identification of the surface arc states extremely difficult. On the other hand, the splitting distance of a pair of WPs connecting with surface arcs is about 0.26 Å (15.5 % of the reciprocal lattice constants) in CuI, which is about ten times larger than that in TaAs [4]. Thus, the ideal feature of Weyl phonons in CuI should be readily detected and is important to the application of topological quantum transport of surface phonons.

It has been established that large and opposite Berry curvature at discrete momentum space can realize valley Hall and phonon Hall effect in electron and phonon systems without inversion symmetry [52,53]. Thus, we can also expect to observe the Weyl phonon Hall effect at the WPs in the CuI phonons. As shown in Figs. 3(c)-(d), applying a temperature gradient along the longitudinal (*xx*) direction, the Weyl phonons excited by circularly polarized light at WPs will be deflected to the transverse (*xy*) direction, depending on the photon polarization. As a result, a temperature difference along the transverse direction would show up since $k_{ph}^{xy} \propto -A \cdot \nabla T_{xx} \cdot \Omega^z - B \cdot \frac{1}{\nabla T_{xx}} \cdot \Omega^z$ (A and B are two system parameters) [54]. The Weyl phonon Hall effect should have advantageous transport applications for the designed phonon dissipation.

As mentioned above, the upper and lower part of Weyl cone belongs to $E_1$ and $B_1$ mode, respectively. The symmetry analysis shows that $B_1$ mode is neither infrared (IR)-active nor Raman (RA)-active, while $E_1$ mode is both IR-active and RA-active. Thus, bulk phonons can be measured by Raman scattering or infrared spectroscopy. On the other hand, the topological edge modes can be detected by the surface sensitive probes such as inelastic x-ray scattering or high resolution electron energy loss spectroscopy [55]. We show the surface phonon density of states (DOS) $\mathcal{G}(\omega)$ at the surface $\bar{\Gamma}$, $\bar{M}$ and $\bar{K}$ point in Fig. S5. It is clear that the peaks of the surface phonon DOS at the $\bar{K}$ point change dramatically and are well separated from the broad bulk continuum, while at the $\bar{M}$ and $\bar{\Gamma}$ point the surface phonon DOS shows a steady tendency. The results reveal that $\bar{K}$ point is the best place to identify the topological edge modes in momentum space.

Since the Weyl points are induced by accidental degeneracy in momentum space, the appearance of Weyl phase is sensitive to perturbations. We found from Fig. S2 that Weyl points would have a global phonon gap under 2% tensile strain. Furthermore, the polariton effect of LO-TO splitting and spin-orbit coupling effect were also discussed (part II Supplementary Material). As shown in Fig. S4, Weyl phase generated accidentally in CuI is not obtained in CuX (X=Cl, Br).

In conclusion, we have theoretically shown that wurtzite CuI features ideal type-II Weyl phonons. The six pairs of accidentally degenerated WPs associated with each other by crystal symmetry and time reversal symmetry have the minimal hybridization between the Weyl nodes and the bulk bands, allowing for readily distinguishable surface states on the Cu-terminated surface and clean open surface arcs on the I-terminated surface. The Raman active and infrared-active $E_1$ mode can help to identify novel Weyl states. We propose that the two-fold degeneracy of Weyl phonons with the opposite Berry curvature in the $k_z = 0.0$ plane can host the Weyl phonon Hall effect promising for the Weyl phononics [56,57]. The proposed properties can be readily detected by light-helicity-resolved Raman spectroscopy connecting the photon helicity and phonon chirality [58-60].

## Acknowledgments

This work was supported by the National Key Research and Development Program of China (Grants No. 2016YFA0300902 and No. 2016YFA0202300), National Basic Research Program of China (Grant No. 2015CB921001), National Natural Science Foundation of China (Grants No. 11774396 and No. 91850120), and "Strategic Priority Research Program (B)" of CAS (Grants No. XDB30000000 and No. XDB07030100).

## References

[1] L. X. Yang, Z. K. Liu, Y. Sun, H. Peng, H. F. Yang, T. Zhang, B. Zhou, Y. Zhang, Y. F. Guo, M. Rahn, D. Prabhakaran, Z. Hussain, S.-K. Mo, C. Felser, B. Yan, and Y. L. Chen, Nat. Phys. **11**, 728 (2015).

[2] B. Q. Lv, H. M. Weng, B. B. Fu, X. P. Wang, H. Miao, J. Ma, P. Richard, X. C. Huang, L. X. Zhao, G. F. Chen, Z. Fang, X. Dai, T. Qian, and H. Ding, Phys. Rev. X **5**, 031013 (2015).

[3] S. M. Huang, Y. S. Xu, I. Belopolski, C. C. Lee, G. Chang, B. Wang, N. Alidoust, G. Bian, M. Neupane, C. Zhang, S. Jia, A. Bansil, H. Lin, M. Z. Hasan, Nat. Commun. **6**, 7373 (2015).

[4] H. Weng, C. Fang, Z. Fang, B. A. Bernevig, X. Dai, Phys. Rev. X **5**, 011029 (2015).

[5] F. Arnold, C. Shekhar, S.-C. Wu, Y. Sun, M. Schmidt, N. Kumar, A. G. Grushin, J. H. Bardarson, R. D. d. Reis, and M. Naumann *et al*., Nat. Commun. **7**, 11615 (2016).

[6] T.-R. Chang, S.-Y. Xu, G. Chang, C.-C. Lee, S.-M. Huang, B. Wang, G. Bian, H. Zheng, D. S. Sanchez, and I. Belopolski *et al*., Nat. Commun. **7**, 10639 (2016).

[7] Y. Chen, Y. Xie, S. A. Yang, H. Pan, F. Zhang, M. L. Cohen, S. Zhang, Nano Lett. **15**, 6974-6978 (2018).

[8] D. Di Sante, P. Barone, A. Stroppa, K. F. Garrity, D. Vanderbilt, and S. Picozzi, Phys. Rev. Lett. **117**, 076401 (2016).

[9] H. Gao, Y. Kim, J. W. F. Venderbos, C. L. Kane, E. J. Mele, A. M. Rappe, W. Ren, Phys. Rev. Lett. **121**, 106404 (2018).

[10] J. Jiang, Z. K. Liu, Y. Sun, H. F. Yang, C. R. Rajamathi, Y. P. Qi, L. X. Yang, C. Chen, H. Peng, C. C. Hwang, S. Z. Sun, S.-K. Mo, I. Vobornik, J. Fujii, S. S. P. Parkin, C. Felser, Binghai Yan, and Y. L. Chen, Nat. Commun. **8**, 13973 (2017).

[11] Z. Wang, M. G. Vergniory, S. Kushwaha, M. Hirschberger, E. V. Chulkov, A. Ernst, N. P. Ong, R. J. Cava, and B. A. Bernevig, Phys. Rev. Lett. **117**, 236401 (2016).

[12] R. Yu, Q. Wu, Z. Fang, H. Weng, Phys. Rev. Lett. **119**, 036401 (2017).

[13] B. Yang, Q. Guo, B. Tremain, R. Liu, L. E. Barr, Q. Yan, W. Gao, H. Liu, Y. Xiang, J. Chen, C. Fang, A. Hibbins, L. Lu, and S. Zhang, Science **359**, 1013 (2018).

[14] W. Gao, B. Yang, M. Lawrence, F. Fang, B. Beri, S. Zhang, Nat. Commun. **7**, 12435 (2016).

[15] T. Liu, S. Zheng, H. Dai, D. Yu, and B. Xia, arXiv:1803.04284 (2018).

[16] V. Peri, M. Serra-Garcia, R. Ilan and S. D. Huber, Nat. Phys. **15**, 357 (2019).

[17] Z. Yang and B. Zhang, Phys. Rev. Lett. **117**, 224301 (2016).

[18] T. Zhang, Z. Song, A. Alexandradinata, H. Weng, C. Fang, L. Lu and Z. Fang, Phys. Rev. Lett. **120**, 016401 (2018).

[19] J. Li, Q. Xie, S. Ullah, R. Li, H. Ma, D. Li, Y. Li and X.-Q. Chen, Phys. Rev. B **97**, 054305 (2018).

[20] Q. Xie, J. Li, M. Liu, L. Wang, D. Li, Y. Li, and X. Q. Chen, arXiv:1801.04048 (2018).

[21] N. P. Armitage, E. J. Mele, and A. Vishwanath, Rev. Mod. Phys. **90**, 015001 (2018).

[22] H. Weng, X. Dai, Z. Fang, J. Phys.: Condens. Matter **28**, 303001 (2016).

[23] X. Wan, A. M. Turner, A. Vishwanath, and S. Y. Savrasov, Phys. Rev. B **83**, 205101 (2011).

[24] F. Li, X. Huang, J. Lu, J. Ma and Z. Liu, Nat. Phys. **14**, 30-34 (2017).

[25] A. A. Soluyanov, D. Gresch. Z. Wang, Q. Wu, M. Troyer, X. Dai and B. A. Bernevig, Nature **527**, 495 (2015).

[26] J. Ruan, S. K. Jian, D. Zhang, H. Yao, H. Zhang, S. C. Zhang and D. Xing, Phys. Rev. Lett. **116**, 226801 (2016).

[27] J. Ruan, S. K. Jian, H. Yao, H. Zhang, S. C. Zhang and D. Xing, Nat. Commun. **7**, 11136 (2016).

[28] X. Huang, L. Zhao, Y. Long, P. Wang, D. Chen, Z. Yang, H. Liang, M. Xue, H. Weng, Z. Fang, X. Dai, and G. Chen, Phys. Rev. X **5**, 031023 (2015).

[29] W. Shi, L. Muechler, K. Manna, Y. Zhang, K. Koepernik, R. Car, J. van den Brink, C. Felser, and Y. Sun, Phys. Rev. B **97**, 060406(R) (2018).

[30] Q. Wang, Y. Xu, R. Lou, Z. Liu, M. Li, Y. Huang, D. Shen, H. Weng, S. Wang and H. Lei, Nat. Commun. **9**, 3681 (2018).

[31] A. A. Zyuzin and R. P. Tiwari, Jetp Lett. **103**, 717-722 (2016).

[32] S. Singh, Q. Wu, C. Yue, A. H. Romero and A. A. soluyanov, Phys. Rev. Materials **2**, 114204 (2018).

[33] M. N. Chernodub, A. Cortijo and M. A. H. Vozmediano, Phys. Rev. Lett. **120**, 206601 (2018).

[34] G. Chang, B. Singh, S.-Y. Xu, G. Bian, S.-M. Huang, C.-H. Hsu, I. Belopolski, N. Alidoust, D. S. Sanchez and H. Zheng *et al.*, Phys. Rev. **97**, 041104(R) (2018).

[35] G. Chang, S. Y. Xu, H. Zheng, B. Singh, C. H. Hsu, G. Bian, N. Alidoust, I. Belopolski, D. S. Sanchez, S. Zhang, H. Liu and M. Z. Hasan, Sci. Rep. **6**, 38839 (2016).

[36] Y. Du, X. Bo, D. Wang, E.-J. Kan, C.-G. Duan, S. Y. Savrasov, and X. Wan, Phys. Rev. B **96**, 235152 (2017).

[37] K. Koepernik, D. Kasinathan, D. V. Efremov, S. Khim, S. Borisenko, B. Büchner, and J. van den Brink, Phys. Rev. B **93**, 201101(R) (2016).

[38] L. Li, H.-H. Xie, J.-S. Zhao, X.-X. Liu, J.-B. Deng, X.-R. Hu and X.-M. Tao, Phys. Rev. B **96**, 024106 (2017).

[39] Y. Sun, S.-C. Wu, M. N. Ali, C. Felser and B. Yan, Phys. Rev. B **92**, 161107(R)

(2015).

[40] A. Tamai, Q. S. Wu, I. Cucchi, F. Y. Bruno, S. Riccò, T. K. Kim, M. Hoesch, C. Barreteau, E. Giannini, C. Besnard, A. A. Soluyanov and F. Baumberger, Phys. Rev. X **6**, 031021 (2016).

[41] L.-L. Wang, N. H. Jo, Y. Wu, Q. S. Wu, A. Kaminski, P. C. Canfield and D. D. Johnson, Phys, Rev. B **95**, 165114 (2017).

[42] Y. Xu, C. Yue, H. Weng and X. Dai, Phys. Rev. X **7**, 011027 (2017).

[43] Z. Wang, D. Gresch, A. A. Soluyanov, W. Xie, S. Kushwaha, X. Dai, M. Troyer, R. J. Cava and B. A. Bernevig, Phys. Rev. Lett. **117**, 056805 (2016).

[44] S. Miyake, S. Hoshino, and T. Takenaka, J. Phys. Soc. Jpn. **7**, 19 (1952).

[45] M. Grundmann, F.-L. Schein, M. Lorenz, T. Böntgen, J. Lenzner, and H. von Wenckstern, Phys. Status Solidi A **210**, 1671 (2013).

[46] G. Autes, D. Gresch, M. Troyer, A. A. Soluyanov and O. V. Yazyev, Phys. Rev. Lett. **117**, 066402 (2016).

[47] Y. Liu, C. S. Lian, Y. Li, Y. Xu and W. Duan, Phys. Rev. Lett. **119**, 255901 (2017).

[48] H. Yang, Y. Sun, Y. Zhang, W.-J. Shi, S. S. P. Parkin, and B. Yan, New J. Phys. **19**, 015008 (2017).

[49] C.-X. Liu, X.-L. Qi, H.-J. Zhang, X. Dai, Z. Fang and S.-C. Zhang, Phys. Rev. B **82**, 045122 (2010).

[50] J. Fritsch and U. Schröder, Phys. Rep. **309**, 209 (1999).

[51] R. Heid and K. P. Bohnen, Phys. Rep. **387**, 151 (2003)

[52] Di Xiao, Gui-Bin Liu, Wanxiang Feng, Xiaodong Xu, and Wang Yao, Phys. Rev. Lett. **108**, 196802 (2012)

[53] Lifa Zhang and Qian Niu, Phys. Rev. Lett. **115**, 115502 (2015)

[54] Tao Qin, Jianhui Zhou, and Junren Shi, Phys. Rev. B **86**, 104305 (2012)

[55] H. Miao, T. T. Zhang, L. Wang, D. Meyers, A. H. Said, Y. L. Wang, Y. G. Shi, H. M. Weng, Z. Fang and M. P. M. Dean, Phys. Rev. Lett. **121**, 035302 (2018).

[56] D. Shin, H. Hübener, U. De Giovannini, H. Jin, A. Rubio, and N. Park, Nat. Commun. **9**, 638 (2018).

[57] J. Liu, W. Hou, C. Cheng, H. Fu, J. Sun and S. Meng, J. Phys.: Condens. Matter **29**, 255501 (2017)

[58] S. Chen, C. Zheng, M. S. Fuhrer and J. Yan, Nano Lett. **15**, 2526 (2015)

[59] H. Zhu *et al.*, Science **359**, 579 (2018).

[60] M. Gao, W. Zhang, and L. Zhang, Nano Lett. **18**, 4424 (2018).

**Figures:**

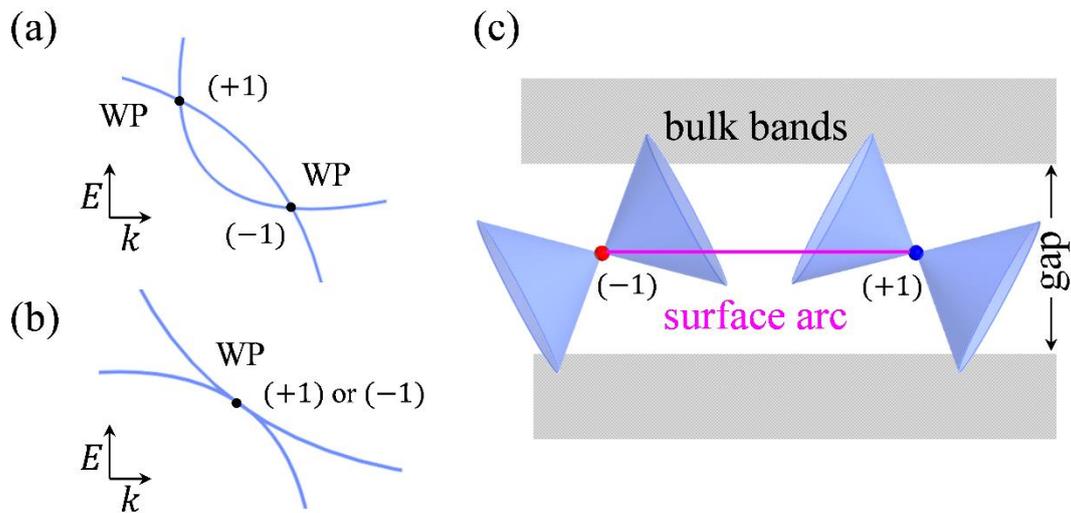

FIG. 1. Schematics of type-II Weyl phase. (a) Two bands cross each other and form two adjacent Weyl points (WPs). (b) Two bands simply touch and form only one WP. (c) Schematics of the ideal type-II Weyl phase. It must meet the requirement in panel (b) that all WPs are related to each other by symmetry. The surface arc connects two WPs (red and blue spheres) with the opposite chirality. The obvious gap away from the Weyl cone minimizes the hybridization between the type-II WPs and the bulk bands.

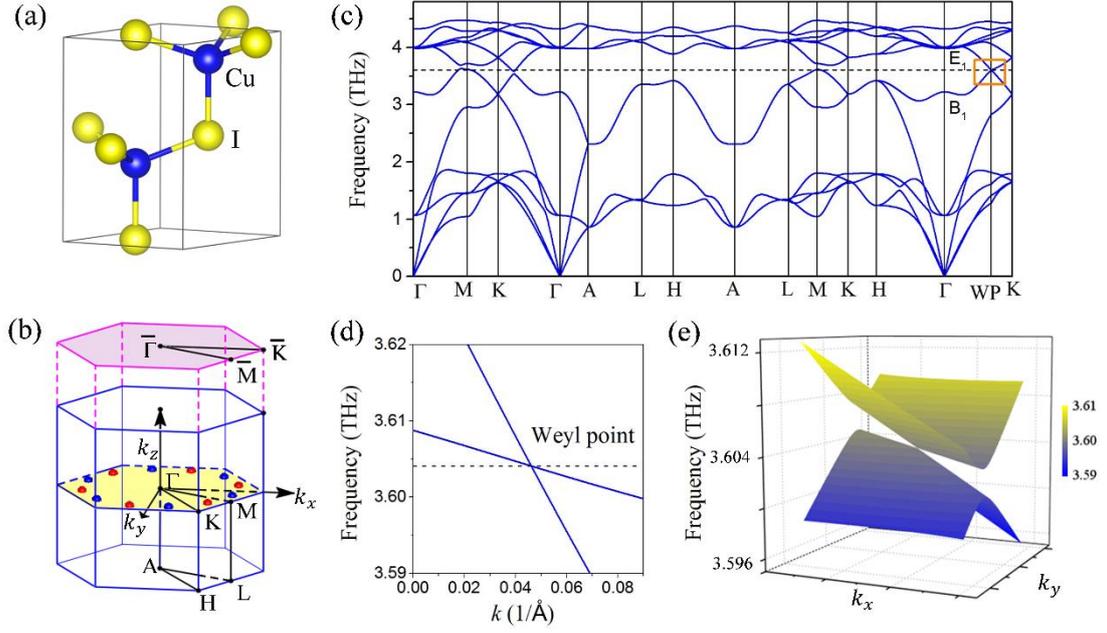

FIG. 2. Type-II Weyl phase in wurtzite CuI. (a) The atomic structures of bulk CuI. Blue and yellow spheres denote the Cu and I atoms, respectively. (b) The bulk Brillouin zone (BZ) and the projected surface BZ for the (0001) surface. The six pairs of type-II Weyl points (WPs) are schematically shown by blue/red spheres denoting the chirality of phonon WP. (c) Phonon dispersion of CuI along high-symmetry momentum path. The phonon WP along the low symmetry momentum path is shown in right section of the phonon dispersion. (d) The enlarged crossing phonon dispersion around one phonon WP along the $\theta = \frac{5\pi}{6}$ direction relative to the $k_x$ axis. (e) Perspective plot of the phonon WP in the $k_z = 0.0$ plane.

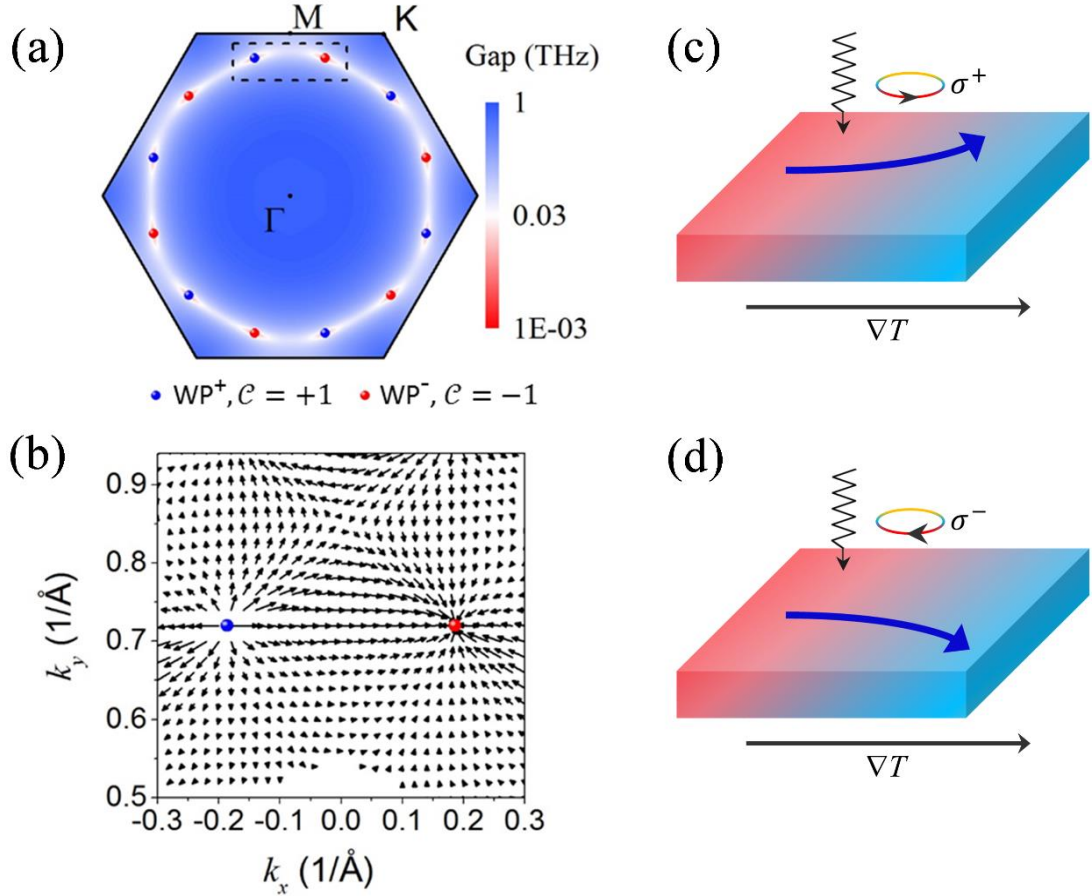

FIG. 3. Topology of Weyl points. (a) The momentum dependent frequency gap between the sixth and seventh phonon band. The chirality of phonon Weyl point (WP) is denoted by blue ($\mathcal{C} = +1$ for WP$^+$) and red ($\mathcal{C} = +1$ for WP$^-$) spheres, respectively. (b) The calculated Berry curvature around a pair of WPs enlarged from the black dashed rectangle in panel (a). The length of each arrow is the magnitude of the in-plane Berry curvature. (c) [(d)] The proposed Weyl phonon Hall effect excited by right-handed (left-handed) circularly polarized light under a temperature gradient. The blue arrows, black arrows and black wave lines denote the Hall current, temperature gradient and light, respectively.

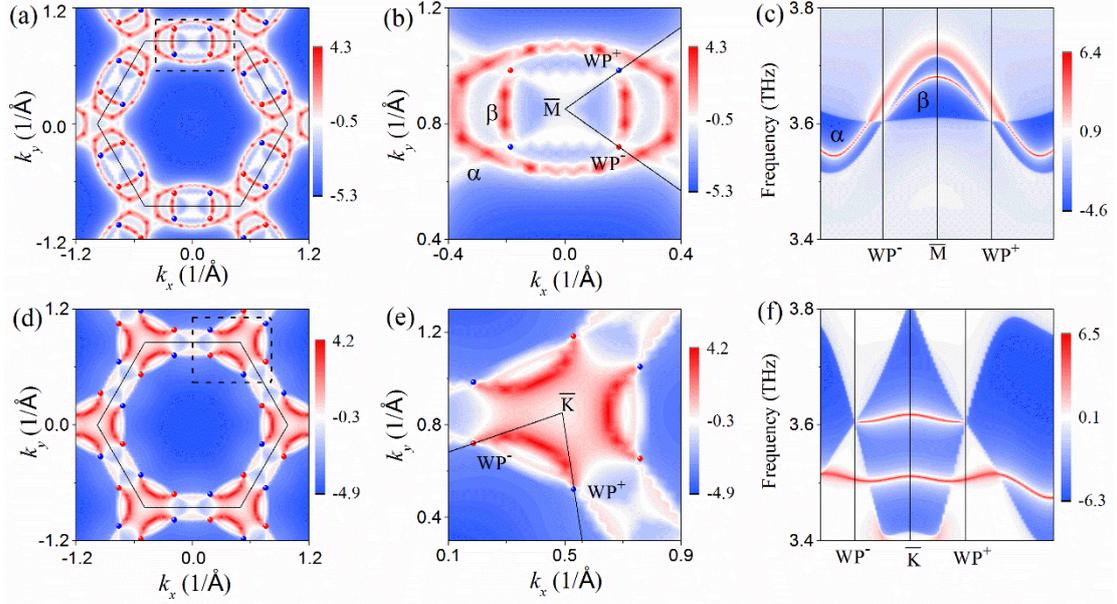

FIG. 4. Surface states and surface arcs. (a) The momentum-resolved surface LDOS for CuI projected onto the top (0001) surface (Cu-terminated surface). (b) The enlarged surface arc of panel (a) signed by black dashed rectangle. It is observed that the open surface arc $\beta$ connects two Weyl points (WPs) with the opposite charges. The length of the surface arc is near 0.26 1/Å. The blue square and red sphere represent the projected WPs with the chirality of +1 and -1, respectively. The state $\alpha$ and $\beta$ are topological trivial and nontrivial, respectively. (c) Surface band structure along the black lines indicated in panel (b). Panels (d)-(f) are the same as panel (a)-(c), respectively, but for the bottom $(000\bar{1})$ surface (I-terminated surface).